\documentclass{PoS}
\title{A nonrelativistic quark model evaluation of exclusive 
$b\to c$ semileptonic decay 
of  triply heavy baryons and
$c\to s,d$ semileptonic decay of  $cb$ baryons}

\ShortTitle{Semileptonic decays of triply and doubly heavy baryons  }

\author{C. Albertus\\
         Departamento de F\'{\i}sica Fundamental e IUFFyM, Universidad de 
	Salamanca, E-37008 Salamanca, Spain\\
        E-mail: \email{albertus@usal.es}}
\author{J.M. Flynn\\
         School of Physics and Astronomy, University of Southampton, Highfield, 
	 Southampton SO17 1BJ, United Kingdom\\
        E-mail: \email{flynn@phys.soton.ac.uk}}
\author{\speaker{E. Hern\'andez}\thanks{   This research was supported by DGI and FEDER funds, under contracts
   FIS2011-28853-C02-02, FPA2010-21750-C02-02, and the Spanish
  Consolider-Ingenio 2010 Programme CPAN (CSD2007-00042),  by Generalitat
  Valenciana under contract PROMETEO/20090090 and by the EU
  HadronPhysics3 Grant no. 283286. C. A. thanks a Juan de 
  la Cierva contract from the
Spanish  Ministerio de Educaci\'on y Ciencia.
}\\
        Departamento de F\'{\i}sica Fundamental e IUFFyM, Universidad de 
	Salamanca, E-37008 Salamanca, Spain\\
        E-mail: \email{gajatee@usal.es}}
\author{J. Nieves\\
         Instituto de F\'{\i}sica Corpuscular (IFIC), Centro Mixto
	 CSIC-Universidad de Valencia, Institutos de Investigaci\'on de Paterna,
	 Apartado 22085, E-46071 Valencia, Spain\\
        E-mail: \email{jmnieves@ific.uv.es}}

\abstract{We present  results for  exclusive $b\to c$ semileptonic decays 
of ground state 
triply-heavy baryons and for  semileptonic $c\to s,d$ decays of doubly
 heavy ground 
state $cb$ baryons. 
In both cases, we have derived for the
  first time heavy quark spin symmetry  relations for
   the hadronic amplitudes near zero recoil. Though   
   strictly valid in the limit of very large heavy quark 
   masses and near zero recoil, they turn out to be reasonable
    accurate for the whole available phase space in these decays and for the
    actual heavy quark masses we use. 
    With these  relations we have made approximate, but model 
    independent, predictions for ratios of decay widths. 
    In the case of spin-1/2 $cb$ baryons, we find that hyperfine mixing
    in the wave function has a great impact on their $c\to s,d$ decay widths. }

\FullConference{Xth Quark Confinement and the Hadron Spectrum,\\
		October 8-12, 2012\\
		TUM Campus Garching, Munich, Germany}

\begin{document}

\section{Introduction}

In this contribution we report on two recent calculations we have done
concerning exclusive $b\to c$ semileptonic decay of triply heavy baryons~\cite{triply}
and $c\to s,d$ semileptonic decay of doubly heavy $cb$ baryons~\cite{cbl}.

The analysis of triply heavy baryons  allows 
to study  the interaction among heavy quarks
in an environment free of valence light quarks. 
With no experimental information available on these systems so far, previous
studies have concentrated on their spectrum
~\cite{Hasenfratz:1980ka,Bjorken,silvestre96,Martynenko:2007je,Roberts:2007ni,Zhang:2009re,Meinel:2010pw}. However, it is 
 likely that triply heavy baryons
would be discovered at LHC~\cite{Chen:2011mb} so that the study of their
properties beyond spectroscopy seems timely.
As for exclusive semileptonic
$c\to s,d$ decays of doubly heavy ground state $cb$ baryons, previous
studies~\cite{sanchis95,Faessler:2001mr,Kiselev:2001fw}  are very limited. This
is in contrast to their corresponding $b\to c$ driven
decays which have been more extensively studied~\cite{sanchis95,Ebert:2004ck,
Hernandez:2007qv,pervin2,Faessler:2009xn,Albertus:2009ww}. 
However, the analysis of the $c\to s,d$ decays of
$cb$ baryons could also give relevant information on heavy quark physics
complementary to the one obtained from the study of their $b\to c$
decays. 

In both calculations we derive for  the first time
heavy quark spin symmetry (HQSS)  approximate expressions for the 
hadronic matrix elements. From these we predict approximate, but model
independent, relations among different decay widths. 

The calculations are done in a nonrelativistic quark model framework. 
We use the AL1 potential  of Refs.~\cite{semay94,silvestre96} which  contains
 $1/r$ and hyperfine terms, that can be understood as originating from a
one-gluon exchange potential, together with a linear confining term. All the
parameters of this potential have been adjusted to de description of light and
heavy meson spectra.

\section{$b\to c$ semileptonic decays of triply heavy baryons}
The wave functions we use to describe triply heavy baryons
have the general form
\begin{eqnarray*}
\Psi_{\alpha_1\alpha_2\alpha_3}=\delta_{f_1h}\delta_{f_2h}
\delta_{f_3h'}\frac{\epsilon_{c_1c_2c_3}}{\sqrt{3!}}\Phi(r_1,r_2,r_{12})
(1/2,1/2,1;s_1,s_2,s_1+s_2)(1,1/2,J;s_1+s_2,s_3,M),
\end{eqnarray*}
where $\alpha_j$ represents the spin (s), flavor (f) and color (c)
quantum numbers of the $j$-th quark. 
 As we are interested only in spin $J=1/2$ or $J=3/2$ ground state baryons,
the total orbital angular momentum is $L=0$. To solve the three-body problem
we shall use a
variational ansatz for the orbital part of
the wave function. We write the
orbital wave functions as the product of three functions,
$\Phi(r_1,r_2,r_{12})=\phi_{hh'}(r_1)\phi_{hh'}(r_2)\phi_{hh}(r_{12})$, 
each one
depending on just one of the three variables $r_1,r_2,r_{12}$, where
$r_1, r_2$  are the
relative distances between quark three and quarks one and two
respectively, and $r_{12}$  is the relative distance between the
first two quarks. 
For each of the $\phi$ functions above we take an expression consisting in the
sum of displaced gaussians of the form
$ \phi(r)=\sum_{j=1}^4a_je^{-b_j^2(r+d_j)^2}$.  We fix the variational parameters by minimizing the
energy while the overall normalization is fixed at the end of the
calculation. 

In Table~\ref{tab:masses} we show the calculated masses of the
triply heavy baryons. Our results agree nicely with the Faddeev evaluation
in Ref.~\cite{silvestre96} using the same interquark potential. For comparison
we also show results obtained in lattice QCD (LQCD)~\cite{Meinel:2010pw}, the
bag model (BM)~\cite{Hasenfratz:1980ka}, relativistic three quark model
(RTQM)~\cite{Martynenko:2007je}, QCD sum rules (QCDSR)~\cite{Zhang:2009re}
and the next to next to leading order calculation using potential
nonrelativistic QCD (NNLO pNRQCD)~\cite{felipe11}. The agreement with the 
LQCD result for the $\Omega^*_{bbb}$ baryon is good. We also agree with the
results in the BM and RTQM calculations. On the other hand the QCDSR results 
are much smaller while the  NNLO pNRQCD calculation predicts larger masses.
\begin{table}
\small{\begin{tabular}{lcc|cccccc}\hline\hline
&This work&\cite{silvestre96}&\cite{Meinel:2010pw}&
  \cite{Hasenfratz:1980ka}   
&\cite{Martynenko:2007je}
     &
  \cite{Zhang:2009re}& \cite{felipe11}  \\
&Variational&Faddeev&LQCD & BM & RTQM  & QCDSR& 
 NNLO pNRQCD\\\hline
$m_{\Omega^*_{bbb}}$&14398&14398& $14371 \pm 12$ & 14300 &    14569  & $13280 \pm 100$& 
$14700\pm300$\\
$m_{\Xi^*_{bbc}}$&11245&--&-- &11200     
& 11287  &$10540 \pm 110$&
$11400\pm300$\\
$m_{\Xi_{bbc}}$&11214&11217&--&--   
 &  11280 &$10300 \pm 100$&
$11400\pm300$\\
$m_{\Xi^*_{ccb}}$&8046&--&--&8030   
  &  8025 &$7450 \pm 160$&
$8150 \pm 300$  \\
$m_{\Xi_{ccb}}$&8018&8019&-- & --  
&  8018& $7410 \pm 130$&
$8150 \pm 300$\\
$m_{\Omega^*_{ccc}}$&4799&4799&--&4790   
  &  4803  & $4670 \pm 150$& 
$4900\pm250$  \\
\hline\hline
\end{tabular}}
\caption{Triply heavy baryon masses (in MeV) obtained with the
AL1 potential of Refs.~\cite{semay94,silvestre96} using our
variational approach. For comparison we also show the results from the
Faddeev calculation performed in Ref.~\cite{silvestre96} using the
same potential.  Predictions within other approaches are also
compiled. The $\Omega^*$ and $\Xi^*$ baryons have total spin 3/2 while the
$\Xi$ ones have total spin 1/2.}
\label{tab:masses}
\end{table}

In the limit of very large heavy quark masses, HQSS predicts for the 
hadronic transition matrix elements near zero recoil~\cite{triply}
\begin{eqnarray*}
\label{eq:transitions}
\Xi_{ccb}\to \Omega^*_{ccc}  &&\hspace{1cm}
 2\eta\, \bar u^{\,\prime\mu} u,
 \label{eq:tran1}\\
\Xi^*_{ccb}\to \Omega^*_{ccc}  &&\hspace{1cm}
-{\sqrt3}\eta\, \bar u^{\,\prime\lambda} \gamma^\mu(1-\gamma_5) u_{\lambda},
  \label{eq:tran2}\\
\Xi_{bbc}\to \Xi_{ccb}  &&\hspace{1cm}
 -\chi \bar u'\big(\gamma^\mu-\frac53\gamma^\mu\gamma_5\big) u,  \label{eq:tran3}\\
\Xi_{bbc}\to \Xi^*_{ccb}  &&\hspace{1cm}
 -\frac2{\sqrt3} \chi \bar u^{\,\prime\mu} u,  \label{eq:tran4}\\
\Xi^*_{bbc}\to \Xi_{ccb}  &&\hspace{1cm}
 -\frac2{\sqrt3} \chi \bar u' u^\mu, \label{eq:tran5} \\
\Xi^*_{bbc}\to \Xi^*_{ccb}  &&\hspace{1cm}
 -2 \chi \bar u'^\lambda \gamma^\mu(1-\gamma_5) u_{\lambda},
 \label{eq:tran6}\\
\Omega^*_{bbb}\to \Xi_{bbc}  &&\hspace{1cm}
 2\xi\, \bar u' u^\mu, \label{eq:tran7}\\
\Omega^*_{bbb}\to \Xi^*_{bbc}  &&\hspace{1cm}
-{\sqrt3}\xi\, \bar u'^\lambda \gamma^\mu(1-\gamma_5) u_{\lambda}, \label{eq:tran8}
\end{eqnarray*}
where the factors $\eta$, $\chi$ and $\xi$ are the Isgur-Wise
functions that depend on the product of four velocities of the two baryons 
$w=v\cdot v'$.  We
evaluate those
Isgur-Wise functions in our model and we  see that, as predicted by the HQSS
relations above,  they reduce to only three
independent ones in very good approximation. With these functions we
get estimates of the $b\to c$ semileptonic
 decay widths that we give in 
 Table~\ref{tab:dw}.
\begin{table}
\begin{center}
\begin{tabular}{cc}\hline\hline
$B\to B'e \bar\nu_e$&\hspace{.5cm}$\Gamma\ [\,{\rm ps}^{-1}]$\hspace*{.5cm}\\\hline
$\Xi_{ccb}\to\Omega^*_{ccc}\,e\bar\nu_e$&$8.01\times 10^{-2}$\\
$\Xi^*_{ccb}\to\Omega^*_{ccc}\,e\bar\nu_e$&$6.28\times 10^{-2}$\\
$\Xi_{bbc}\to\Xi_{ccb}\,e\bar\nu_e$&$7.98\times 10^{-2}$\\
$\Xi_{bbc}\to\Xi^*_{ccb}\,e\bar\nu_e$&$2.42\times 10^{-2}$\\
$\Xi^*_{bbc}\to\Xi_{ccb}\,e\bar\nu_e$&$1.17\times 10^{-2}$\\
$\Xi^*_{bbc}\to\Xi^*_{ccb}\,e\bar\nu_e$&$7.74\times 10^{-2}$\\
$\Omega^*_{bbb}\to\Xi_{bbc}\,e\bar\nu_e$&$3.95\times 10^{-2}$\\
$\Omega^*_{bbb}\to\Xi^*_{bbc}\,e\bar\nu_e$&$6.34\times 10^{-2}$\\
\hline\hline
\end{tabular}
\end{center}
\caption{Estimated decay widths  in units of ${\rm ps}^{-1}$. We use
$|V_{bc}|=0.0410$. }
\label{tab:dw}
\end{table}

As the $d\Gamma/dw$ differential decay width peaks at  
$w$ values very close to 1~\cite{triply}, one can make further approximations
 valid in that region. The lepton tensor is approximately given by
${\cal L}^{\alpha\beta}(q)\approx \frac{-\pi q^2}{6}\,\big(g^{\alpha\beta}-
\frac{q^\alpha q^\beta}{q^2}\big)$, where $q$ is the  total four-momentum
of the leptonic system. Besides in the
 product of lepton and hadron tensors we can
approximate $w\approx 1$ and $\frac{(v\cdot q)^2}{q^2}\approx\frac{(v\cdot q)(v'\cdot q)}{q^2}\approx
\frac{(v'\cdot q)^2}{q^2}$.  Further assuming 
$m_{B_{bbc}}\approx m_{B^*_{bbc}}\ ;\ m_{B_{ccb}}\approx  m_{B^*_{ccb}}$, we
predict the approximate ratios
\begin{eqnarray*}
 \frac{2\Gamma(\Xi^*_{bbc}\to
  \Xi_{ccb})}{\Gamma(\Xi_{bbc}\to \Xi^*_{ccb})}&\approx&
1,\\ \frac{\Gamma(\Xi^*_{bbc}\to\Xi^*_{ccb})}{4\Gamma(\Xi_{bbc}\to\Xi_{ccb})-
  10\Gamma(\Xi_{bbc}\to\Xi^*_{ccb})}&\approx& 1.
\end{eqnarray*}
These approximate, but model independent, predictions are satisfied in our 
own calculation at the 3.4\% and 0.25\% level 
respectively and  we expect them to hold in other approaches as well.

\section{$c\to s,d$ semileptonic decays of doubly heavy $cb$ baryons}
The baryons involved in the present calculation are given in 
Table~\ref{tab:baryons}. The quark model masses quoted
 have been taken from our previous works in
Refs.~\cite{Albertus:2003sx,Albertus:2009ww}, where they were obtained
using the AL1 potential of
Refs.~\cite{semay94,silvestre96}. All the details on the wave functions and how
they are evaluated can be found in Refs.~\cite{cbl,
Albertus:2003sx,Albertus:2006wb}. Experimental masses shown in 
Table~\ref{tab:baryons} are isospin 
 averaged over the values reported by the particle data group~\cite{pdg10}. 
 For the actual
calculation of the decay widths we  use experimental masses 
 whenever possible.
\begin{table}
\begin{center}
\small{\begin{tabular}{ccccccc}\hline\hline
Baryon &~~~~$J^\pi$~~~~&~~~~ $I$~~~~&~~~~$S^\pi$~~~~& 
Quark content &\multicolumn{2}{c}{Mass\ [MeV]}\\\cline{6-7}

       &       &         &   &      & Quark model   & Experiment                
\\       &       &         &   &      & 
\cite{Albertus:2003sx,Albertus:2009ww}   & \cite{pdg10} \\  
\hline
$\Xi_{cb}$ &$\frac12^+$& $\frac12$ &$1^+$&$cbn$&6928&--
\\
$\Xi'_{cb}$ &$\frac12^+$& $\frac12$ &$0^+$&$cbn$&6958&--
\\
$\Xi^*_{cb}$ &$\frac32^+$& $\frac12$ &$1^+$&$cbn$&6996&--
\\
$\Omega_{cb}$  &$\frac12^+$& 0 &$1^+$&$cbs$&7013&--\\
$\Omega'_{cb}$  &$\frac12^+$& 0 &$0^+$&$cbs$&7038&--\\
$\Omega^*_{cb}$  &$\frac32^+$& 0 &$1^+$&$cbs$&7075&--\\

\hline
$\Lambda_b$ &$\frac12^+$& 0 &$0^+$&$udb$&5643&$5620.2\pm1.6$
\\
$\Sigma_b$ &$\frac12^+$& 1 &$1^+$&$nnb$&5851&$5811.5\pm2.4$
\\
$\Sigma^*_b$ &$\frac32^+$& 1 &$1^+$&$nnb$&5882&$5832.7\pm3.1$
\\
$\Xi_b$  &$\frac12^+$&$\frac12$&$0^+$&$nsb$&5808&$5790.5\pm2.7$
\\
$\Xi'_b$  &$\frac12^+$&$\frac12$&$1^+$&$nsb$&5946&--
\\
$\Xi^*_b$ &$\frac32^+$&$\frac12$&$1^+$&$nsb$&5975&--
\\
$\Omega_b$  &$\frac12^+$& 0 &$1^+$&$ssc$&6033&$6071\pm40$
\\
$\Omega^*_b$  &$\frac32^+$& 0 &$1^+$&$ssc$&6063&--
\\\hline
\hline
\end{tabular}}
\end{center}
\caption{Quantum numbers of the baryons involved in this study. 
 The usual
 classification scheme in which the two heavy quarks or the two light quarks
 have well defined total spin is used. $J^\pi$ and $I$ are the spin-parity and
 isospin of the baryon,  $S^\pi$ is the spin-parity of the two
 heavy or the two light quark subsystem. $n$ denotes a $u$ or $d$
 quark. }
\label{tab:baryons}
\end{table}

The classification scheme shown in Table~\ref{tab:baryons} assumes that the two
heavy quarks or the two light quarks have well defined total spin $S$. 
This is not correct for spin-1/2 states. Due to the finite value of the heavy quark
 masses, the hyperfine
interaction between a light  quark and a heavy quark can
admix both $S$=0 and 1 components into the wave function. We neglect these 
effects  for the $\Xi_{b}$ and $\Xi'_{b}$ states as the hyperfine matrix elements
linking the two states are proportional to the inverse of the $m_b$ quark mass.
 On the other hand, for the $\Xi_{cb},\,\Xi'_{cb}$ ($\Omega_{cb},\,\Omega'_{cb}$)
the effect is only suppressed by the $c$ quark mass and it is relevant.
As a result, the actual physical spin-1/2 $cb$ baryons are
admixtures of the $\Xi_{cb},\,\Xi'_{cb}$
($\Omega_{cb},\,\Omega'_{cb}$) states.  The physical states that we obtain in
our model  are given  by~\cite{Albertus:2009ww}\footnote{Note that here we use
 the   order $cb$, whereas in Ref.~\cite{Albertus:2009ww}, we used $bc$. Thus 
 our
  $\Xi'_{cb}$ and $\Omega'_{cb}$ states, where the heavy quark
  subsystem is coupled to spin zero,
  differ in  sign with those used in Ref.~\cite{Albertus:2009ww}.}
\begin{eqnarray*}
\Xi_{cb}^{(1)}&=&-0.902\ \Xi'_{cb}+0.431\ \Xi_{cb}\ \ ;\ M_{\Xi_{cb}^{(1)}}=6967\,{\rm MeV},\nonumber\\
\Xi_{cb}^{(2)}&=&\hspace{.25cm}0.431\  \Xi'_{cb}+0.902\ \Xi_{cb}\ \ ;
\ M_{\Xi_{cb}^{(2)}}=6919\,{\rm MeV},\\
\label{eq:mixedxi}
%
\Omega_{cb}^{(1)}&=&-0.899\  \Omega'_{cb}+0.437\ \Omega_{cb}\ \ ;\ M_{\Omega_{cb}^{(1)}}=7046\,{\rm MeV},\nonumber\\
\Omega_{cb}^{(2)}&=&\hspace{.25cm}0.437\  \Omega'_{cb}+0.899\ \Omega_{cb}\ \ ;
\ M_{\Omega_{cb}^{(2)}}=7005\,{\rm MeV}.
\label{eq:mixedomega}
\end{eqnarray*} 
These physical spin-1/2 $cb$ baryon
states turn out to be very close to the states (here $B$ stands for $\Xi$ or
$\Omega$)
\begin{eqnarray*}
\widehat B_{cb}&=&-\frac{\sqrt3}{2}B'_{cb}+\frac{1}{2}B_{cb},\nonumber\\
\widehat B'_{cb}&=&\frac{1}{2}B'_{cb}+\frac{\sqrt3}{2}B_{cb}.
\label{eq:qchqss}
\end{eqnarray*}
 in which  the $c$ 
and the light $q$ quark
couple to  well defined spin $S_{cq}=1$ ($\widehat B_{cb}$) or 0 ($\widehat B'_{cb}$), and then the $b$
quark couples to that state to make the baryon with total spin 1/2.
Hyperfine mixing for the $\widehat B_{cb},\,\widehat B'_{cb}$ states
is much less important since it is inversely proportional to the $b$
quark mass.

While  masses are not very
sensitive to hyperfine mixing, it was pointed out
in Ref.~\cite{pervin1} that hyperfine mixing could
greatly affect the decay widths of doubly heavy spin-1/2 $cb$ baryons.  This
assertion was confirmed for  $b\to c$ semileptonic decay 
in Refs.~\cite{pervin2,Albertus:2009ww} and for electromagnetic 
transitions in Refs.~\cite{Albertus:2010hi,Branz:2010pq}.  We expected
configuration mixing to also play an important role for $c\to s,d$
semileptonic decay of $cb$ baryons.

The  decay widths we evaluate appear in Tables~\ref{tab:resctos} and
\ref{tab:resctod}. We show our full results and, in between parentheses, the results where 
configuration
mixing is not considered. In all cases we find a good agreement with the few other previous 
calculations. We also see that configuration mixing effects are 
very important for transitions to final states where the two light quarks 
couple to spin 1, where we
find enhancements or reductions as large as a factor of 2.
\begin{table}[h!!!]
\footnotesize{\begin{tabular}{llc}\hline\hline
&\multicolumn{2}{c}{$\Gamma \ [10^{-14}\,{\rm GeV}]$}\\
&{This work}&Others\\
\hline
$\Xi^{(1)\,+}_{cbu}\to\Xi^0_b\, e^+\nu_e$& 3.74 (3.45)&(3.4)\,\cite{sanchis95}\\
$\Xi^{(2)\,+}_{cbu}\to\Xi^0_b\, e^+\nu_e$& 2.65 (2.87)\\
$\Xi^{(1)\,+}_{cbu}\to\Xi'^0_b\, e^+\nu_e$& 3.88
(1.66)&$2.44\div3.28^\dagger$\,\cite{Faessler:2001mr}\\
$\Xi^{(2)\,+}_{cbu}\to\Xi'^0_b\, e^+\nu_e$&1.95 (3.91)\\
$\Xi^{(1)\,+}_{cbu}\to\Xi^{*\,0}_b\, e^+\nu_e$& 1.52 (3.45)\\
$\Xi^{(2)\,+}_{cbu}\to\Xi^{*\,0}_b\, e^+\nu_e$& 2.67 (1.02)\\
$\Xi^{(2)\,+}_{cbu}\to\Xi^0_b\, e^+\nu_e+\Xi'^0_b\, e^+\nu_e+
\Xi^{*\,0}_b\, e^+\nu_e$& 7.27 (7.80)
&$(9.7\pm1.3)^\ddag$\,\cite{Kiselev:2001fw}\\
$\Xi^{*\,+}_{cbu}\to\Xi^{0}_b\, e^+\nu_e$&  4.08\\
$\Xi^{*\,+}_{cbu}\to\Xi'^{0}_b\, e^+\nu_e$&0.747\\
$\Xi^{*\,+}_{cbu}\to\Xi^{*\,0}_b\, e^+\nu_e$& 5.03\\\hline\hline
\end{tabular}\hspace{.5cm}
\begin{tabular}{ll}\hline\hline
&{\hspace*{.375cm}$\Gamma \ [10^{-14}\,{\rm GeV}]$}\\
\hline
$\Omega^{(1)\,0}_{cbs}\to\Omega^-_b\, e^+\nu_e$&\hspace*{.5cm} 7.21 (3.12)\\
$\Omega^{(2)\,0}_{cbs}\to\Omega^-_b\, e^+\nu_e$&\hspace*{.5cm} 3.49 (7.12)\\
$\Omega^{(1)\,0}_{cbs}\to\Omega^{*\,-}_b\, e^+\nu_e$&\hspace*{.5cm} 2.98 (6.90)\\
$\Omega^{(2)\,0}_{cbs}\to\Omega^{*\,-}_b\, e^+\nu_e$&\hspace*{.5cm} 5.50 (2.07)\\
$\Omega^{*\,0}_{cbs}\to\Omega^-_b\, e^+\nu_e$&\hspace*{.5cm} 1.35\\
$\Omega^{*\,0}_{cbs}\to\Omega^{*\,-}_b\, e^+\nu_e$&\hspace*{.5cm} 10.2\\\hline\hline
\end{tabular}}
\caption{$\Gamma$ decay widths for  $c\to s$ decays.  We use $|V_{cs}|=0.973$. Results where configuration
mixing is not considered are shown in between parentheses. The result with a
$\dag$ corresponds to the decay of the $\widehat \Xi_{cb}$
state. The result 
with an ${\ddag}$ is our
estimate from the total decay width and the branching ratio
given in~\cite{Kiselev:2001fw}. } 
\label{tab:resctos}
\end{table}
\begin{table}[h!!!]
\small{\begin{tabular}{ll}\hline\hline
&{\hspace*{.5cm}$\Gamma \ [10^{-14}\,{\rm GeV}]$}\\
\hline
$\Xi^{(1)\,+}_{cbu}\to\Lambda^0_b\, e^+\nu_e$&\hspace*{.5cm} 0.219 (0.196)\\
$\Xi^{(2)\,+}_{cbu}\to\Lambda^0_b\, e^+\nu_e$&\hspace*{.5cm}  0.136 (0.154)\\
$\Xi^{(1)\,+}_{cbu}\to\Sigma^0_b\, e^+\nu_e$&\hspace*{.5cm}  0.198 (0.0814)\\
$\Xi^{(2)\,+}_{cbu}\to\Sigma^0_b\, e^+\nu_e$ &\hspace*{.5cm} 0.110 (0.217)\\
$\Xi^{(1)\,+}_{cbu}\to\Sigma^{*\,0}_b\, e^+\nu_e$&\hspace*{.5cm}  0.0807 (0.184)\\
$\Xi^{(2)\,+}_{cbu}\to\Sigma^{*\,0}_b\, e^+\nu_e$&\hspace*{.5cm}  0.147 (0.0556)\\
$\Xi^{*\,+}_{cbu}\to\Lambda^{0}_b\, e^+\nu_e$&\hspace*{.5cm}   0.235\\
$\Xi^{*\,+}_{cbu}\to\Sigma^{0}_b\, e^+\nu_e$&\hspace*{.5cm}  0.0399\\
$\Xi^{*\,+}_{cbu}\to\Sigma^{*\,0}_b\, e^+\nu_e$&\hspace*{.5cm}  0.246\\\hline\hline
\end{tabular}\hspace{2cm}
\begin{tabular}{ll}\hline\hline
&{\hspace*{.5cm}$\Gamma \ [10^{-14}\,{\rm GeV}]$}\\
\hline
$\Omega^{(1)\,0}_{cbs}\to\Xi^-_b\, e^+\nu_e$&\hspace*{.5cm}  0.179 (0.164)\\
$\Omega^{(2)\,0}_{cbs}\to\Xi^-_b\, e^+\nu_e$&\hspace*{.5cm}  0.120 (0.133)\\
$\Omega^{(1)\,0}_{cbs}\to\Xi'^-_b\, e^+\nu_e$&\hspace*{.5cm}  0.169 (0.0702)\\
$\Omega^{(2)\,0}_{cbs}\to\Xi'^-_b\, e^+\nu_e$&\hspace*{.5cm}  0.0908 (0.182)\\
$\Omega^{(1)\,0}_{cbs}\to\Xi^{*\,-}_b\, e^+\nu_e$&\hspace*{.5cm}  0.0690 (0.160)\\
$\Omega^{(2)\,0}_{cbs}\to\Xi^{*\,-}_b\, e^+\nu_e$&\hspace*{.5cm}  0.130 (0.0487)\\
$\Omega^{*\,0}_{cbs}\to\Xi^-_b\, e^+\nu_e$&\hspace*{.5cm}  0.196\\
$\Omega^{*\,0}_{cbs}\to\Xi'^-_b\, e^+\nu_e$&\hspace*{.5cm}  0.0336\\
$\Omega^{*\,0}_{cbs}\to\Xi^{*\,-}_b\, e^+\nu_e$&\hspace*{.5cm}  0.223\\\hline\hline
\end{tabular}}\caption{$\Gamma$ decay widths for  
$c\to d$ decays.  We  use $|V_{cd}|=0.225$. In between parentheses
we show the results without configuration mixing.}
\label{tab:resctod}
\end{table}

Now, in the limit of very large heavy quark masses we can use  HQSS to
 approximately evaluate the hadronic matrix 
elements for semileptonic transitions between hatted states 
($S_{cq}$ well defined). Close to zero recoil those matrix elements are given by~\cite{cbl}
\begin{itemize}
\item $\widehat B_{cb}\to\Lambda_b,\Xi_b$ \hspace{.5cm}$\frac1{\sqrt3}\eta\,
\bar u'\big(-\gamma^\mu\gamma_5\big)u$,
\item $\widehat B'_{cb}\to \Lambda_b,\Xi_b$\hspace{.5cm}
$\eta\,\bar u'\gamma^\mu u$,
\item $\widehat B^*_{cb}\to \Lambda_b,\Xi_b$\hspace{.5cm}
$-\eta\,\bar u^{\prime\,} u^\mu$,
\item  $\ \widehat B_{cb}\to \Sigma_b, \Xi'_b, \Omega_b$\hspace{.5cm}
$\beta\bar u'(\gamma^\mu-\frac23\gamma^\mu\gamma_5)u$,
\item $\ \widehat B'_{cb}\to \Sigma_b, \Xi'_b, \Omega_b$\hspace{.5cm}
$\frac1{\sqrt3}\beta\,
\bar u'(-\gamma^\mu\gamma_5)u$,
\item $\ \widehat B^*_{cb}\to \Sigma_b, \Xi'_b, \Omega_b$\hspace{.5cm}
$\frac{1}{\sqrt3}\beta\,\bar u'u^\mu$,
\item $\ \widehat B_{cb}\to \Sigma^*_b, \Xi^*_b, \Omega^*_b$\hspace{.5cm}
$\frac{1}{\sqrt3}\beta\,
\bar u^{\prime\mu}u$,
\item $\ \widehat B'_{cb}\to \Sigma^*_b, \Xi^*_b, \Omega^*_b$\hspace{.5cm}
$-\beta\bar u^{\prime\,\mu}u$,
\item $\ \widehat B^*_{cb}\to \Sigma^*_b, \Xi^*_b, \Omega^*_b$\hspace{.5cm}
$-\beta\,\bar u^{\prime\,\lambda}\gamma^\mu(1-\gamma_5)u_\lambda$.
\end{itemize}
These relations impose restrictions on the form factors for the different 
decays  that are  well satisfied
within our model~\cite{cbl} over the whole $w$ range accessible for the decays
and for the actual 
heavy quark masses. 
With the use of the above HQSS relations, and the approximations (exact a zero
recoil) similar to the ones described in the previous section, we are able to predict 
approximate, but model independent, relations among decay widths for hatted 
states. Those are given in the following where we also show the results of our
full calculation using physical  (close to hatted) states.
\begin{eqnarray*}
&&\Gamma(\widehat\Xi_{cb}\to\Lambda_b)\approx\Gamma(\widehat\Xi^*_{cb}\to\Lambda_b)
\hspace{1cm}
 0.219 \approx 0.235,\\
&&\Gamma(\widehat B_{cb}\to\Xi_b)\approx\Gamma(\widehat B^*_{cb}\to\Xi_b)
\hspace{1cm}
    0.179 \approx 0.196\ (B=\Omega)\hspace{1cm} 3.73\approx 4.08,\ (B=\Xi)
\end{eqnarray*}
\begin{eqnarray*}
\Gamma(\widehat\Xi'_{cb}\to\Sigma_b)\approx3\Gamma(\widehat\Xi^*_{cb}\to\Sigma_b)
&\approx&\frac32\Gamma(\widehat\Xi_{cb}\to\Sigma^*_b)\approx
\frac12\Gamma(\widehat\Xi'_{cb}\to\Sigma^*_b)\\
0.110\approx   0.120&\approx&0.121\approx0.074,\\\\
\Gamma(\widehat B'_{cb}\to\Xi'_b)\approx3\Gamma(\widehat B^*_{cb}\to\Xi'_b)
&\approx&\frac32\Gamma(\widehat B_{cb}\to\Xi^*_b)\approx
\frac12\Gamma(\widehat B'_{cb}\to\Xi^*_b)\nonumber\\
   0.097   \approx   0.101&\approx&0.104\approx0.065\ (B=\Omega)\nonumber\\
   1.95   \approx   2.24&\approx&2.29\approx1.34\ (B=\Xi),\\\\
\Gamma(\widehat\Omega'_{cb}\to\Omega_b)\approx3\Gamma(\widehat\Omega^*_{cb}\to\Omega_b)
&\approx&\frac32\Gamma(\widehat\Omega_{cb}\to\Omega^*_b)\approx
\frac12\Gamma(\widehat\Omega'_{cb}\to\Omega^*_b)\\
   3.49   \approx                                   
   4.05&\approx&4.48\approx2.75,
\end{eqnarray*} 
\begin{eqnarray*}
&&\Gamma(\widehat\Xi^*_{cb}\to\Sigma^*_b)\approx
\Gamma(\widehat\Xi^*_{cb}\to\Sigma_b)+\Gamma(\widehat \Xi_{cb}\to\Sigma_b)
\hspace{.5cm}
 0.246 \approx 0.238,\\
&&\Gamma(\widehat B^*_{cb}\to\Xi^*_b)\approx
\Gamma(\widehat B^*_{cb}\to\Xi'_b)+\Gamma(\widehat B_{cb}\to\Xi'_b)\hspace{.5cm}
 0.223 \approx 0.203\ (B=\Omega)\hspace{.5cm}
 5.03 \approx 4.62\ (B=\Xi),\\
&&\Gamma(\widehat\Omega^*_{cb}\to\Omega^*_b)\approx
\Gamma(\widehat\Omega^*_{cb}\to\Omega_b)+
\Gamma(\widehat\Omega_{cb}\to\Omega_b)\hspace{1cm}
 10.2 \approx 8.56,
\end{eqnarray*}
Our results agree with the HQSS based 
predictions at the 10\% level in most cases. The large 
discrepancies present in a few notable cases are mainly due
to the different phase space as a result of baryon mass differences~\cite{cbl}. 
We expect the above relations 
to hold in other approaches to the same
level of accuracy.


\begin{thebibliography}{99}

\bibitem{triply} J.M. Flynn, E. Hern\'andez and J. Nieves, Phys. Rev. D {\bf 85}, 014012
(2012).

\bibitem{cbl} C. Albertus, E. Hern\'andez and J. Nieves, Phys. Rev. D {\bf 85}, 
094035 (2012).

\bibitem{Hasenfratz:1980ka}
  P.~Hasenfratz, R.~R.~Horgan, J.~Kuti, J.~M.~Richard,
  Phys.\ Lett.\  {\bf B94}, 401 (1980).
  
\bibitem{Bjorken} J.D. Bjorken, Preprint FERMILAB-Conf-85/69.

\bibitem{silvestre96} B. Silvestre-Brac, Few-Body Systems  {\bf 20}, 1 (1996).

\bibitem{Martynenko:2007je}
  A.~P.~Martynenko,
  Phys.\ Lett.\  {\bf B663}, 317 (2008).

\bibitem{Roberts:2007ni}
  W.~Roberts and M.~Pervin,
  Int.\ J.\ Mod.\ Phys.\  A {\bf 23}, 2817 (2008).

\bibitem{Zhang:2009re}
  J.~R.~Zhang and M.~Q.~Huang,
  Phys.\ Lett.\   {\bf B674} (2009) 28.

\bibitem{Meinel:2010pw}
  S.~Meinel,
  Phys.\ Rev.\  D{\bf 82}, 114514 (2010).



\bibitem{Chen:2011mb}
  Y.~-Q.~Chen, S.~-Z.~Wu, JHEP {\bf 1108}, 144 (2011);
  Erratum-ibid.\  {\bf 1109}, 089 (2011).

\bibitem{sanchis95} M. A. Sanchis-Lozano, Nucl. Phys.  B {\bf 440}, 251 (1995).

\bibitem{Faessler:2001mr}
  A.~Faessler et al.,
  Phys.\ Lett.\   {\bf B518}, 55 (2001).

\bibitem{Kiselev:2001fw}
  V.~V.~Kiselev and A.~K.~Likhoded,
  Phys.\ Usp.\  {\bf 45}, 455 (2002)
  [Usp.\ Fiz.\ Nauk {\bf 172}, 497 (2002)].


\bibitem{Ebert:2004ck} 
  D.~Ebert, R.~N.~Faustov, V.~O.~Galkin and A.~P.~Martynenko,
  Phys.\ Rev.\ D {\bf 70}, 014018 (2004);
  Erratum-ibid.\ D {\bf 77}, 079903 (2008).

\bibitem{Hernandez:2007qv}
  E.~Hernandez, J.~Nieves, J.~M.~Verde-Velasco,
  Phys.\ Lett.\   {\bf B663}, 234   (2008).
%
\bibitem{pervin2}  W. Roberts and M. Pervin,
Int. J. Mod. Phys. A  {\bf 24}, 2401 (2009).
%
\bibitem{Faessler:2009xn} 
  A.~Faessler et al,
  Phys.\ Rev.\ D {\bf 80}, 034025 (2009).

\bibitem{Albertus:2009ww}
  C.~Albertus, E.~Hernandez and J.~Nieves,
  Phys.\ Lett.\   {\bf B683}, 21 (2010).

\bibitem{semay94} C. Semay, and B. Silvestre-Brac, Z. Phys. C {\bf 61}, 271 (1994).


\bibitem{felipe11}
    F.J. Llanes-Estrada, O.I. Pavlova, R. Williams,   
    Eur. Phys. J.  C {\bf 72}, 2019 (2012).

\bibitem{Albertus:2003sx}
  C.~Albertus, J.~E.~Amaro, E.~Hernandez and J.~Nieves,
  Nucl.\ Phys.\  A {\bf 740}, 333 (2004).


\bibitem{Albertus:2006wb}
  C.~Albertus, E.~Hernandez, J.~Nieves and J.~M.~Verde-Velasco,
  Eur.\ Phys.\ J.\  A {\bf 31}, 691 (2007); Erratum-ibid.  A {\bf 36},
  119  (2008).


\bibitem{pdg10} K. Nakamura et al. (Particle Data Group), J. Phys. G {\bf 37}, 
075021 (2010).

\bibitem{pervin1}  W. Roberts and M. Pervin,
Int. J. Mod. Phys. A {\bf 23}, 2817 (2008).


\bibitem{Albertus:2010hi}
  C.~Albertus, E.~Hernandez, J.~Nieves,
 Phys.\ Lett.\   {\bf B690},  265 (2010).

\bibitem{Branz:2010pq} 
  T.~Branz et al.,
  Phys.\ Rev.\ D {\bf 81}, 114036 (2010).






\end{thebibliography}
\end{document}